\documentclass{article}
\usepackage{float}
\usepackage{amsmath}
\usepackage{amssymb}
\usepackage{graphicx}
\usepackage{esint}

\usepackage{algorithm}
\usepackage[noend]{algpseudocode}
\usepackage{hyperref}
\hypersetup{
    colorlinks=false,
}


\date{}

\begin{document}
\title{Vector Field-based Simulation of Tree-Like Non-Stationary Geostatistical Models}


\author{Viviana Lorena Vargas, Sinesio Pesco\\
  \small Pontificia Universidade Catolica, Rio de Janeiro\\
  \small vivagra@puc-rio.br, sinesio@puc-rio.br\\
}

\maketitle


\begin{abstract}
In this work, a new non-stationary multiple point geostatistical algorithm called vector field-based simulation is proposed. The motivation behind this work is the modeling of a certain structures that exhibit directional features with branching, like a tree, as can be frequently found in fan deltas or turbidity channels. From an image construction approach, the main idea of this work is that instead of using the training image as a source of patterns, it may be used to create a new object called a training vector field (TVF). This object assigns a vector to each point in the reservoir within the training image. The vector represents the direction in which the reservoir develops. The TVF is defined as an approximation of the tangent line at each point in the contour curve of the reservoir. This vector field has a great potential to better capture the non-stationary nature of the training image since the vector not only gives information about the point where it was defined but naturally captures the local trend near that point.
\end{abstract}

\section{Introduction}
\label{sec:1}
Geostatistics is a set of tools that model and predict spatial-temporal natural phenomena. Measurements are expensive and time consuming, so the phenomena is usually characterized by taking only some samples at different locations. Geostatistical methods provide estimates for locations where samples were not taken as well as the uncertainty of that prediction.

Multiple-point geostatistics methods (MPS) are geostatistical techniques developed to overcome difficulties that appeared in traditional methods; these include variogram-based and object-based methods. The variogram approaches are based on two-point statistics and therefore do not have the capacity to reproduce complex patterns such as those present in meandering channels. Some examples of these methods are presented in \cite{jour}, \cite{jode}, \cite{tyler}, \cite{chile}. The object methods consist of the geometry parametrization and geometry placement. The first is the description of the object by parametric geometries, which means the object can be characterized by simple geometries depending on the parameters. These parameters (for example sinuosity, depth, length and width) must be properly chosen according to the available information such as seismic data and outcrops.  The geometry placement, refers to the disposition of the objects created on the grid according to a probability distribution. They are randomly placed until an established criterion is reached. Examples of this kind of simulation are in \cite{hala}, \cite{detran}, \cite{haska}.

MPS can account for correlations among more than two positions at a time, meaning these kind of methods may reproduce the connectivity of many locations. MPS depends on the concept of a training image, which was introduced in  \cite{SR} and \cite{GS}; it is a source of spatial continuity that represents the prior geological knowledge of the modeled phenomena. The training image is a conceptual image that is assumed to contain all possible structures of interest believed to appear in the phenomena. Multiple-point methods present different approaches, for example, using the training image to reproduce statistics and conditional probabilities \cite{S}, \cite{VRIES} or to reproduce patterns; instances of these approaches are presented in \cite{AC}, \cite{ZSJ}, \cite{HC2}.

An important problem in geostatistics involves providing methods for modeling structures whose image representation does not correspond to the typical realization of a stationary point process. In fact, natural processes frequently have a complex nature that makes stationary modeling inadequate, so these processes can be better understood by non-stationary models.

MPS methods have been used to create non-stationary methods as proposed in \cite{AR}, \cite{AC}, \cite{HC}. In \cite{AR}, the training image is partitioned into regions where it is assumed that a stationary model can be applied, and only one direction is associated with each area. This direction is used to rotate a new training image; the rotated image is then used in a stationary simulation at the corresponding region. In \cite{HC2}, Gabor filters are used to automatically divide the training image into stationary zones, and a stationary simulation is applied in each zone. There are more methods using MPS for non-stationary modeling; for a detailed description, see \cite{MC}.

\begin{figure}[h!]
		\centering
		\includegraphics[width=\linewidth]{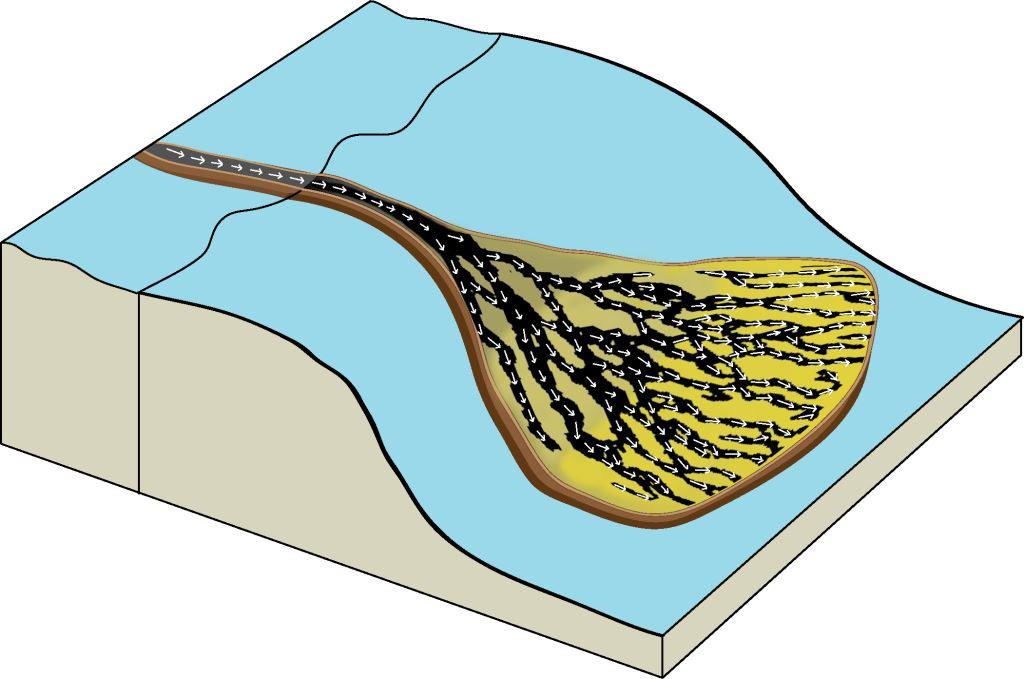}
		\caption{Vector field defined on the training image}
	\label{im_sin}
	\end{figure}

In contrast with the previous methods, where directions are assigned to regions, this work will construct a vector field giving a vector to each point within the reservoir, which is illustrated in Fig. \ref{im_sin}.

The interest of this work is the modeling of a specific set of geological phenomena with a branching geometrical structure that frequently can be found  in fan deltas, alluvial fans and turbidites channels in submarine fans for which a stationary model is not appropriate. In this paper, we present a new non-stationary multiple-point simulation for geologic structures, called vector field simulation (VECSIM), whose spatial continuity can be assumed to be reflected by training images similar to the ones illustrated in Fig. \ref{im_sin}. These images present a branching structure, similar to a tree, that will be called a tree-like geometry. The geometry is characterized by the orientation and width of the channels varying throughout the image, much like the branches of a tree. The main feature of a tree-like image is that at each point, one vector can be perceived in which direction the image appears to be developing. The main novelty in the proposed algorithm is the definition of a vector field to the training image, as illustrated in Fig. \ref{im_sin}, which represents the flow followed at each point. This field is used to control the MPS simulation so that the realizations obtained have a similar directional orientation to that of the training image. The main contributions of this work are to present a method for extracting the vector field from the training image and a new multiple point simulation method for modeling turbidite channels.

The first section will be about the structures modeled and the characteristics of the tree-like images. In the following section, the vector field for these kinds of images is obtained; this is done by defining the vectors on the contours of the image that are generated by successive erosions. The use of the contour to define the vector field is another novelty in the methodology. Then, the vector field-based simulation is presented. Finally, several examples to show the applicability of the method are presented.

\section{Conceptualization of the model}
\label{sec:2}
Turbidite deposits are generated by turbidity currents and related gravity flows. These currents are caused by catastrophic events such as storm surges, shocks induced by earthquakes, and failures of sediments due to steep slopes or river floods. They move huge amounts of sand, mud and pebbles forming turbidites in oceans, and lakes either in shallow or deep waters. Turbidite reservoirs are distinguished by a complex structure of sand bodies arranged in channels and lobes deposited in deep water environments. This kind of reservoir still represents an important source of deep water oil exploration and the characterization of these systems are very important in the petroleum industry.

\begin{figure}[h!]
\centering
\includegraphics[width=\linewidth]{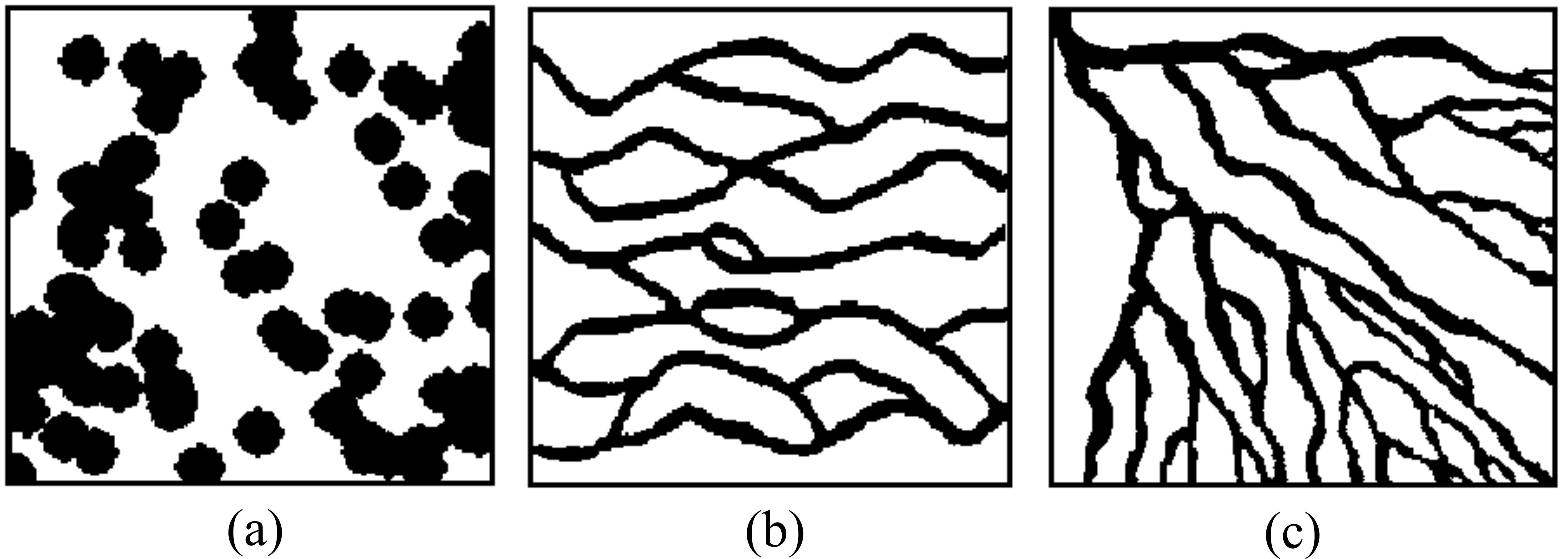}
\caption{Comparison between 3 images to illustrate the notion of tree-like image}\label{comp}
\end{figure}

\begin{figure}[h!]
		\centering
		\includegraphics[width=8.5cm]{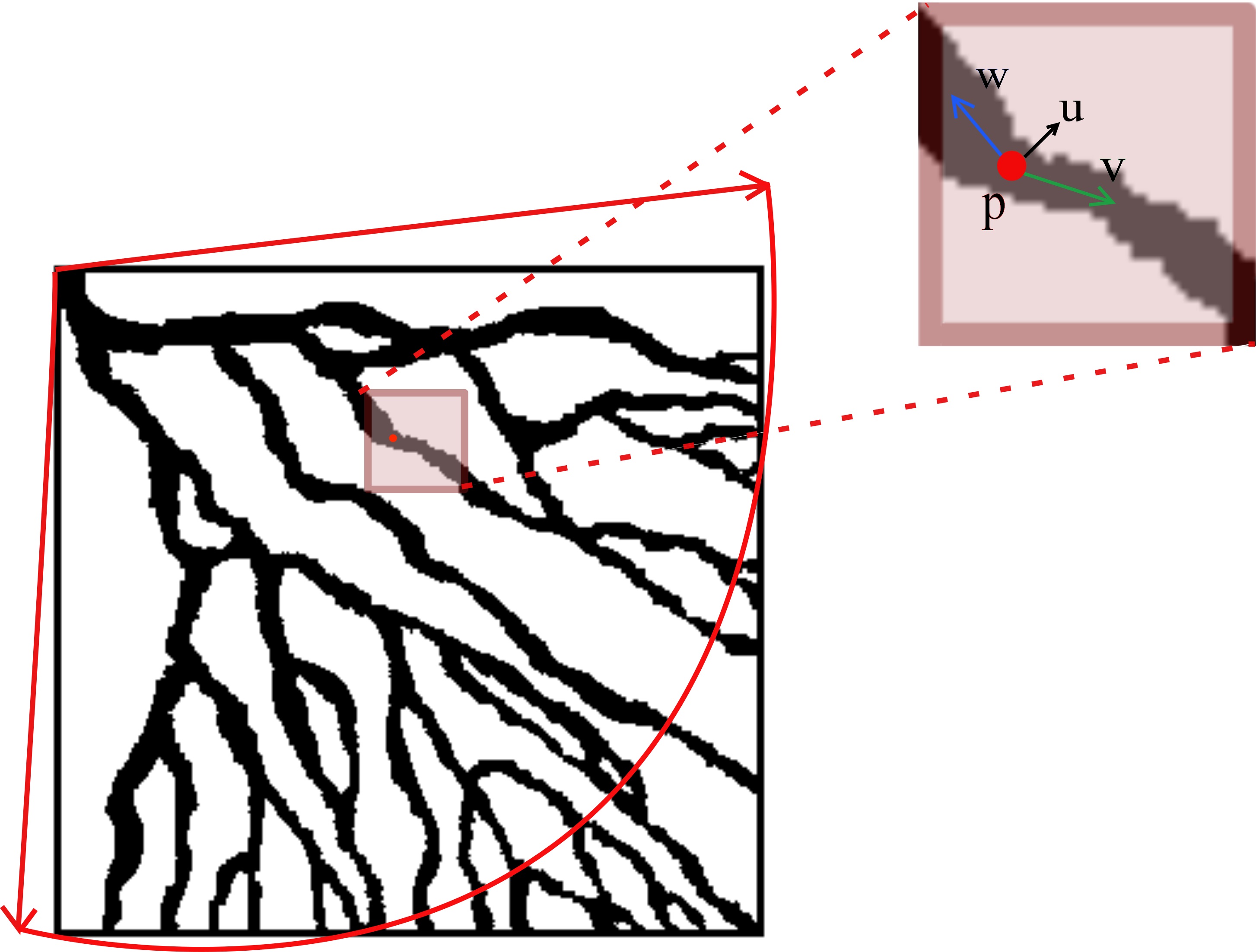}
		\caption{If the point $p$ is only observed locally as in the small square, it is not possible to decide what direction is the appropriate ($w$ or $v$). However, globally, the behavior of the structure reflects the appropriate direction $v$}\label{pref_dir}
		
	\end{figure}

This phenomenon occurs on a surface, and we seek to describe it through 2D models that can be interpreted as its projection on the horizontal plane. Because of the specific structure of the modeled object, we restrict our methodology to images with a special geometry, which we call ``tree-like''. There is not a formal definition for this term, but tree-like images should have the following features: 

\begin{enumerate}
\item At each black point, one direction in which the channel appears to be developing can be perceived.  Figure \ref{comp}(a) is not tree-like because if a black point is taken, there is no a distinguishable direction in which the object is being formed. Both images in Fig. \ref{comp}(b and c) meet this condition.


\item A directional interval, \textbf{DI}, with diameter less than $\pi$ can be defined in the image. DI is defined as an interval containing the directions that are present in the image. It can be established visually; for example, in Fig. \ref{pref_dir} the directions in the tree-like geometry are inside the red cone. Then, the directional interval is determined by the opening angle of the cone and its position. 

\end{enumerate}

If a point $p$ is taken in a tree-like image, tree vectors stand out (see Fig. \ref{pref_dir}). In the direction of the vector $u$ the point moves away from the object in a short distance, because in this direction the channel is crossed through its width. In directions of the vectors $w$ and $v$ the point $p$ tends to continue for a longer distance within the object. Observing the point $p$ with respect to the region in the blue square (Fig. \ref{pref_dir}), these two vectors show the tendency of the channel to develop. However, one of them goes against the flow of the channel. To decide which one does so, the region must be seen as part of the entire channel system.  According to the general structure of the image, $v$ can be chosen as the appropriate vector in the point $p$ because the system seems to develop in that direction. In practice, this is done by the use of the directional interval, since the direction of the vector $v$ is in the interval, contrary to the vector $w$.

Both images \ref{comp}(b and c) meet the two conditions, and our proposed method could be applied to both; however, the images that will be used in our examples are of the type of (c) because they represent the kind of phenomena that have been modeled. The image has a starting region from which branches are born, ramify and continue; this characteristic gives them the appearance of having a tree-like structure. 

\section{Methodology}
\label{}

The multiple-point method presented follows a pattern-based approach of image construction. The basic idea in these methods is to extract a pattern base from the training image with a fixed template and to use it in a stochastic simulation, where the measure of similarity between patterns is based on the values contained in the image. Our idea is to change the training image by including vector field data extracted from the modeled structure.

The method is divided into two parts. In the first, due to the directional characteristics given by the tree-like training images, a vector field is obtained defining a vector at each point within the reservoir. In the second part, this field is employed to guide the simulation algorithm.

\subsection{Training vector field}

From the training image \textbf{TI}, $ti$ is defined as the function that assigns to $p=(i,j)$ the value of \textbf{TI} at the point $p$. Since this work is restricted to binary images, the function $ti$ is given by

	\begin{equation*}
		ti(p) = 
		\begin{cases}
			1, & \text{ if }  \textbf{TI}  \text{ contains sand in } $p$ \\
			0, & \text{ if }  \textbf{TI}  \text{ contains no sand in } $p$\\
		\end{cases}
		\end{equation*}

This training image characterization shows whether one point is inside the channel system or not. Additionally, one vector is assigned to each point in the reservoir. Based on these vectors, a new object called \textbf{trai\-ning vector field (TVF)} is defined, which contains the vectors over the reservoir in the training image, thus representing the flow of the reservoir. The \textbf{TVF} is represented by the function $tvf$ given by

	\begin{equation*}
			tvf(p) = 
				\begin{cases}
						 v_p, & \text{ if }  ti(p) = 1 \\
						\text{ND},& \text{ if }  ti(p) = 0 \\
				\end{cases}
	\end{equation*} 
	
Where $v_p$ is a vector defined for points within the reservoir whose direction represents how the channel is developing at the point $p$. At points outside of the channel system, there is not a defined vector; this is denoted by \textbf{ND}.

It is assumed that the image contour is a good source of information about its development. For this reason we propose it as a fundamental tool in the definition of the TVF. First, the image is decomposed into a sequence of contour curves. Then, the vector at each point is defined according to the contour curve to which it belongs.


\subsubsection{Image decomposition}
\label{decomposition}

To extract the contours, successive erosions to the training image are applied. Erosion is a morphological operator that takes two inputs: the image to be eroded and the structuring element (also known as a kernel), which determines the effect of the erosion on the image. The basic effect is to erode away the boundaries of regions of foreground pixels (black pixels). Thus areas of foreground pixels shrink in size, and holes within those areas become larger.

\begin{figure}[h!]
				
				\centering
				\includegraphics[width=\linewidth]{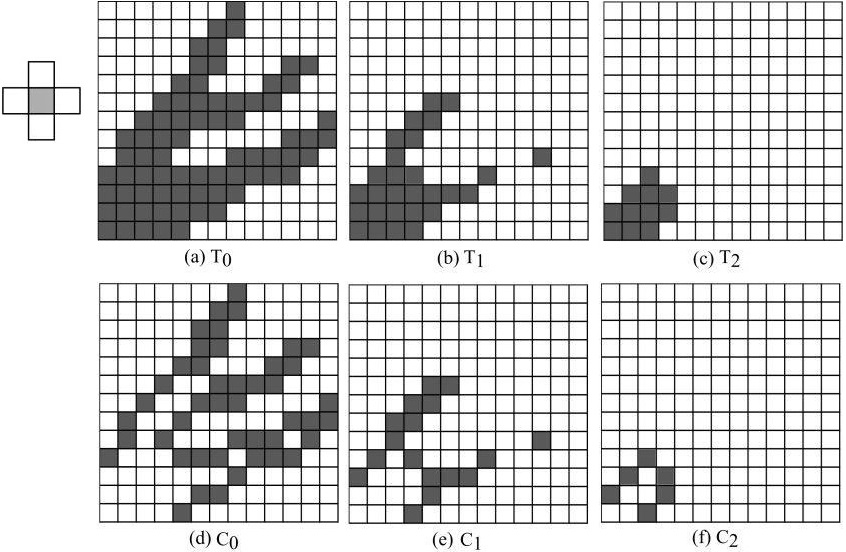}
				\caption{Illustration of two successive erosions $T_1$ and $T_2$  from the image $T_0$, using the structuring element on the left. The images $C_n$ are the contours of the $T_n$}\label{seq}
				\end{figure}

Figure \ref{seq} illustrates the erosion applied to the training image in (a) using the structuring element on the left.  $T_1$ is the result of the erosion of the training image $T_0$, which is formed by the black points in $T_0$ that are not on its contour $C_0$, so $T_1 = T_0 - C_0$. The erosion is repeated successively and two sequences are generated, one with the contours $C_0, C_1, \ldots, C_k$ and another with the erosions $T_0, T_1, \ldots, T_k$. Since $T_{n+1} = T_n-C_n$ is strictly smaller than $T_{n}$, the body in the sequence $T_{n}$ is becoming increasingly smaller at each step. The training image $T_{0}$ is described as the following union

\begin{equation} \label{union}
T_0 = C_0 \cup C_1 \cup \ldots C_{k-1} \cup T_k
\end{equation}

in the explicit example given by Fig. \ref{seq}, $T_0 = C_0 \cup C_1 \cup T_2$.


Multiple criteria can be used to stop the erosion process; for example, when the number of points in the eroded image $T_n$ is less than a certain rate of the total points in the initial image. In addition, the number of connected components can increase at each erosion, and then other alternative is to observe the number of connected components of the image $T_n$, and based on that, decide when to stop. 

\subsubsection{Vector field}\label{contour}

\begin{figure}[h!]
		
			\centering
			\includegraphics[width=7.5cm]{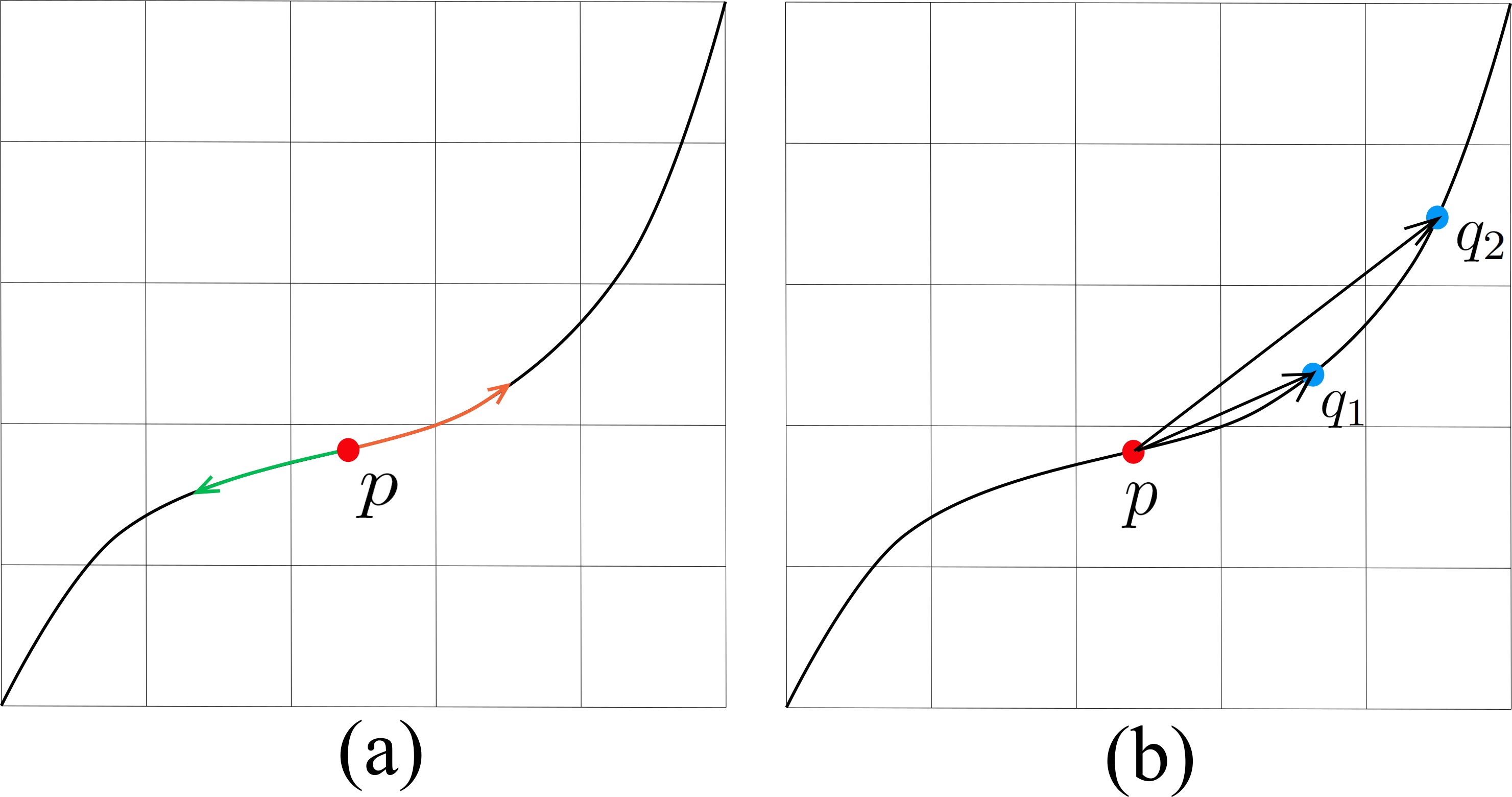}
			\caption{(a) Two possible paths in the contour. (b) Two secants passing through points $p$ and $q_1$; $p$ and $q_2$}\label{dir_contorno}
		\end{figure}

The field at each point is defined using the contour to which it belongs. Since the contour is a curve, at each point there are two vectors that the point can follow to continue on the contour, see Fig. \ref{dir_contorno}(a). The vector within the directional interval is chosen as the value of the vector field at the point. To define a vector at the point $p$, the tangent line to the curve at the point is found. The secant lines passing through p and nearby points $q$ are considered, as in Fig. \ref{dir_contorno}(b). The tangent is obtained at the limit when the point $q$ tends towards $p$. In our discrete case, the tangent will be approximated using two secants. 
 


To define the secants, a path on the contour is followed in the direction given by the directional interval. To take a step of size 1 from the point $p$ on the contour means to choose one neighbor point $q$ ($q$ is immediately next to $p$), which is on the contour and meets the established ID, i.e., the direction of the segment $\overline{pq}$ is inside the directional interval. If more than one point satisfies the previous condition, then one of them is randomly selected. A step of size $n$ means to successively take $n$ steps of size 1.

To construct the TVF, two step sizes $n$ and $m$ are fixed. By taking these steps from a point $p$, two final positions are reached: points $ q_n $ and  $ q_m $. The direction $\alpha_i $ of the secant passing through $p$ and $q_i$ is defined as the angle between the line $\overline{pq_i}$ and $x$ axis, for $i = n,m$. The direction of the vector at point $p$ is defined as the average of these two directions. The norm is defined as one.

\begin{figure}[h!]
	
		\centering
		\includegraphics[width= 5.0cm]{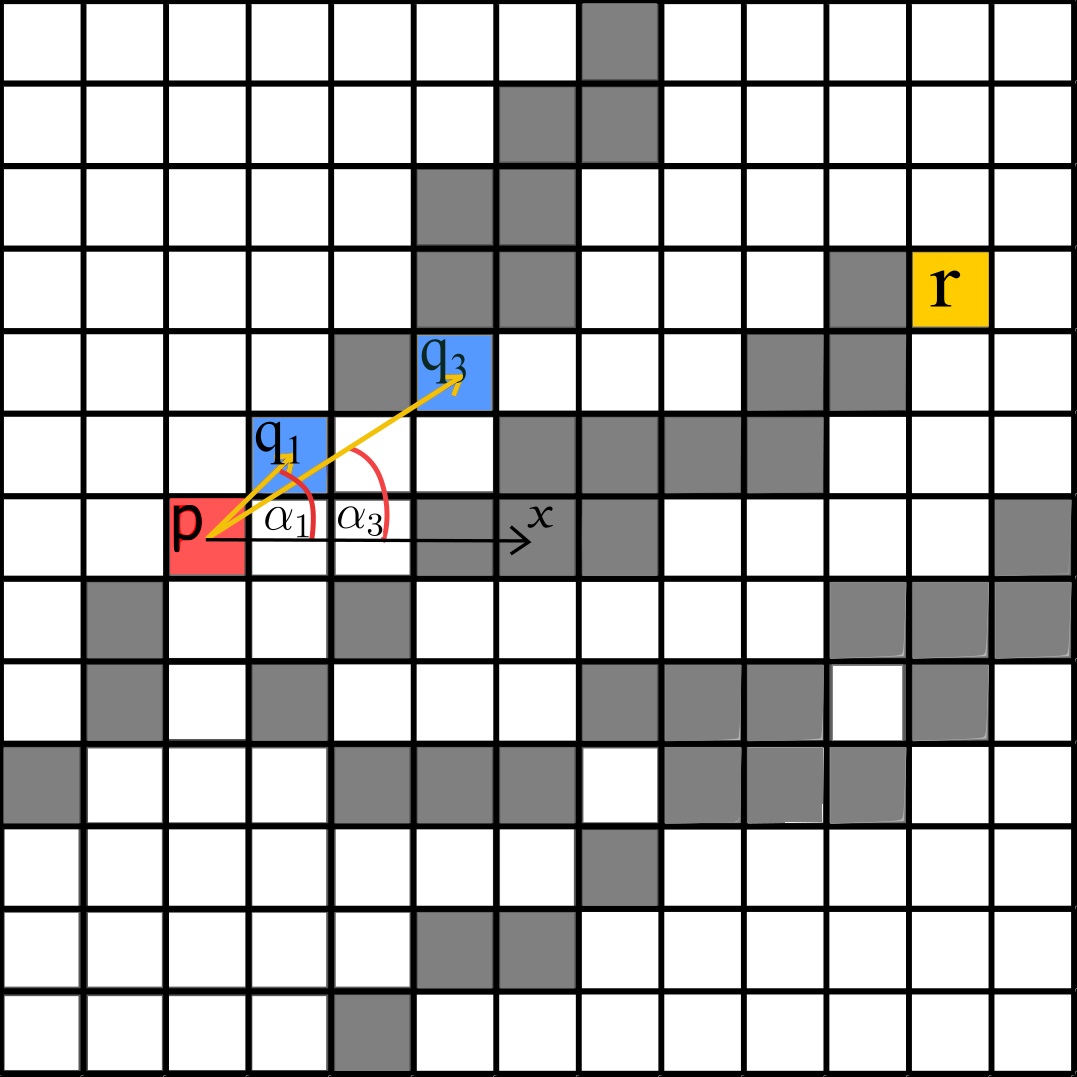}
		\caption{Steps of size 1 and 3 from point $p$}\label{steps}
		\end{figure}

In Fig. \ref{steps}, the two points $q_1$ and $q_3$ take steps of size 1 and 3 respectively, from the point p. The arrows indicate the direction; $\alpha_1$ and $\alpha_3$ are the respective angles formed with the $x$ axis. For points shown as $r$, no steps can be taken from them; in this case the definition of the vector is reserved for later and will be based on the previously defined vectors.


\begin{figure}[t!]
	   
		\centering
		\includegraphics[width=9.5cm]{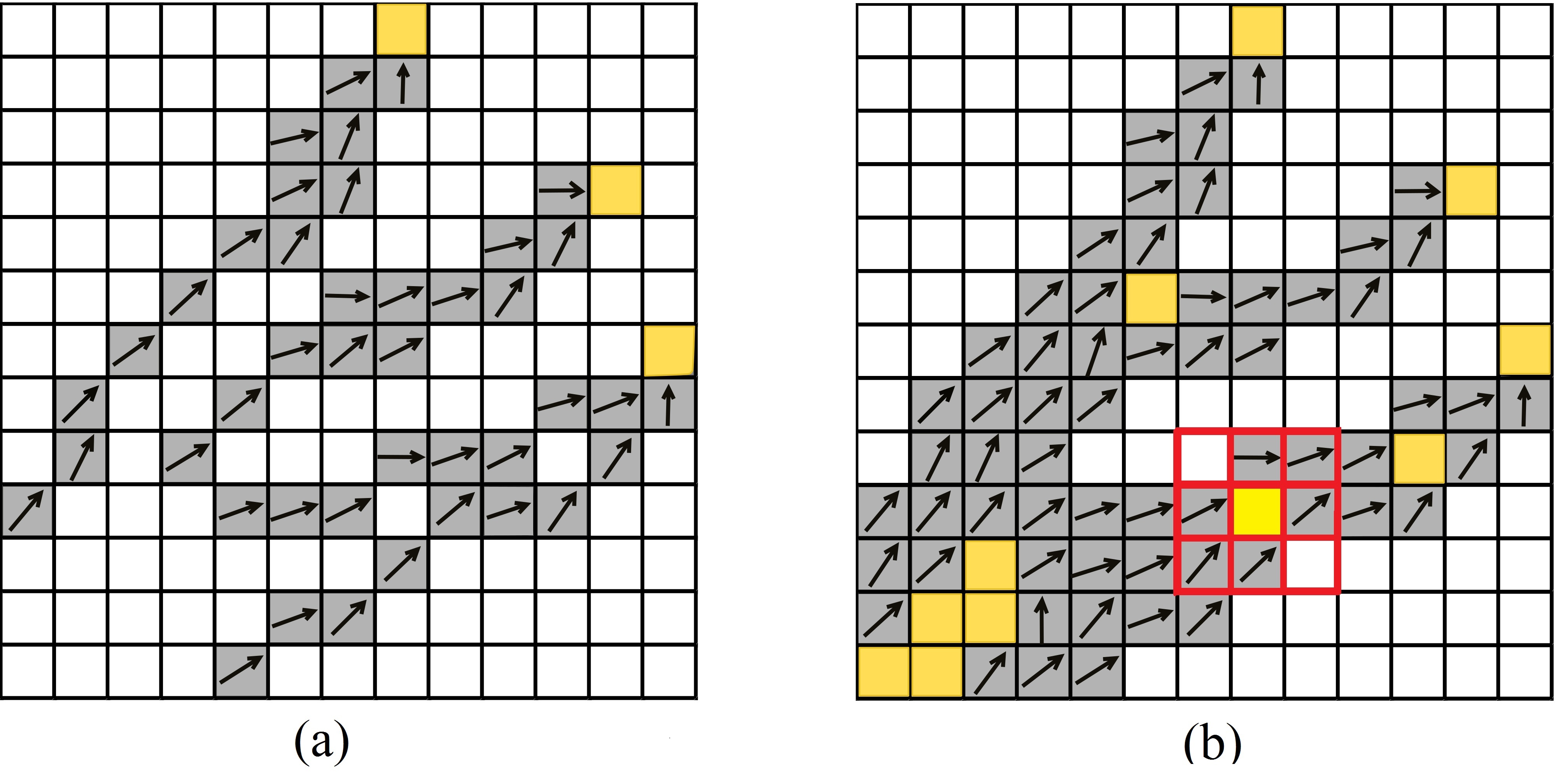}
		\caption{(a) (a) The TVF defined in the first element $C_0$ of the sequence $C_n$. (b) TVF defined in the contours $C_0, C_1, C_2$ and the yellow points do not have an assigned direction yet}\label{TVF}
		\end{figure}

In Fig. \ref{TVF}(a), the set of vectors in the first element $ C_0 $ of the contour sequence is illustrated. Yellow positions have no following point and thus do not have an established vector yet.

The TVF is defined in the points of the successive contours $C_n$. However, the vectors must still be assigned to the interior points, $ T_k $, and those that belong to some $C_n$ but do not have a vector yet. For that points the direction of the TVF is defined by the average of the known directions within a convenient neighborhood (usually to one or two points of distance). This process continues until every point has a defined direction.

In Fig. \ref{TVF}(b), the TVF in the successive contours $C_0, C_1, C_2$, given in Fig. \ref{seq} (d, e, f) is shown. The yellow points represent the points in $T_3 = T_2 - C_2$ together with the points in the contours that do not have a defined vector. A template is centered in these points and the direction of the vector is defined as the average of the known directions inside the template. In Fig. \ref{TVF}(b) an example of the template for that interpolation is shown in red. Finally, in Fig. \ref{vantagem}, the entire TVF is shown.

The previous method to find the TVF can be summarized in algorithm  \ref{alg_TVF}.

\makeatletter
\def\BState{\State\hskip-\ALG@thistlm}
\makeatother

\begin{algorithm}[h!]
\caption{TVF generation}\label{alg_TVF}
\begin{algorithmic}[1]
\State Generate the contour $C_0$ of the training image $T_0$.
\State A new image $T_1 = T_0 - C_0$ is obtained.
\State Keep repeating (1), (2) to get $C_i$ and $T_{i+1}$ for each step $i$. 
\State Find the vector at each contour point in the sequence $C_n$, using steps of size $ n $ and $ m $. 
\State For all points where no vector has been defined, the vector is found by interpolation of known neighbor values.

\end{algorithmic}
\end{algorithm}

\section{Vector field synthesis}

The stationary modeling method is not the most appropriate for the modeled phenomenon where the characteristics vary through the image. To address this non-stationarity, unlike other pattern-based methods, VECSIM uses the training vector field instead of the training image as a source of patterns.



\begin{figure}[h!]
		\centering
		\includegraphics[width=8.0cm]{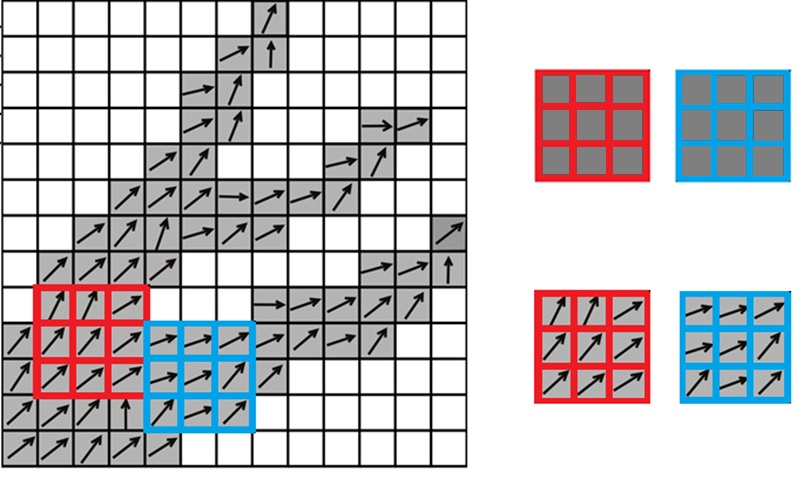}
		\caption{Illustration of two patterns extracted using only the training image, as well as the same patterns from the training vector field}\label{vantagem}
	
		\end{figure}

The advantage of using patterns coming from a vector field can be better appreciated in Fig. \ref{vantagem}. Considering only the training image, the patterns in blue and red are identical. However, they belong to two different branches of the image with different directions. The vector field pattern seeks to preserve local trends to globally reproduce the structure of the tree.

\begin{figure}[h!]
		\centering
		\includegraphics[width=\linewidth]{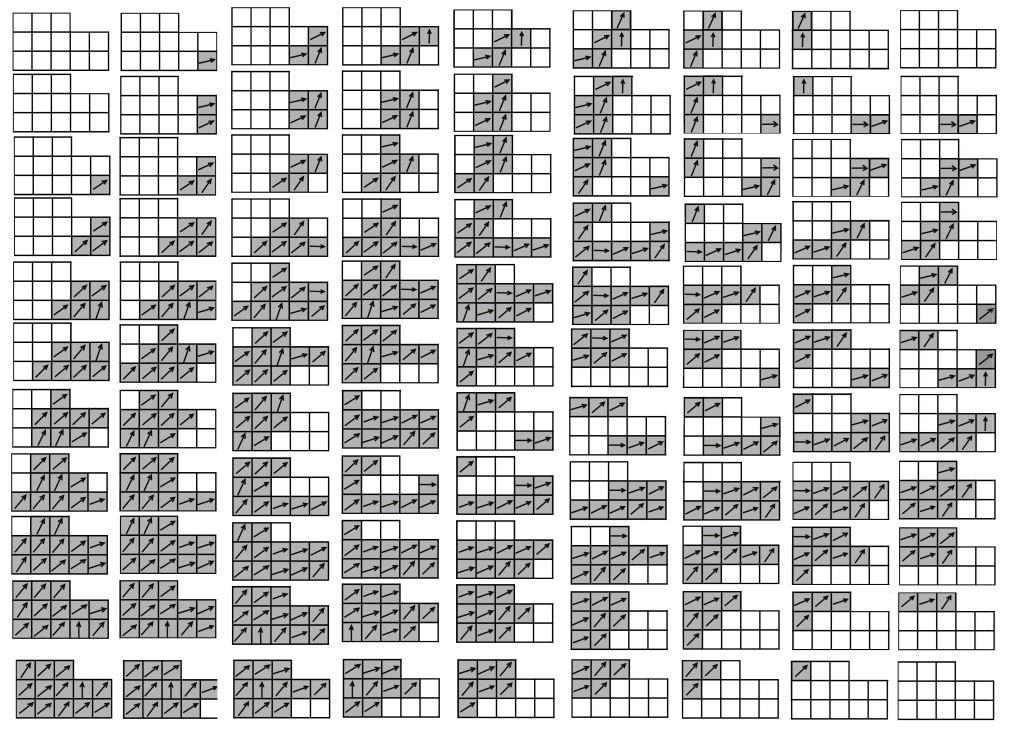}
		\caption{Pattern base of the training vector field in Fig. \ref{vantagem} using an L-template}\label{PB}
		\end{figure}

\begin{figure}[h!]
	  		\centering
		\includegraphics[width=7.5cm]{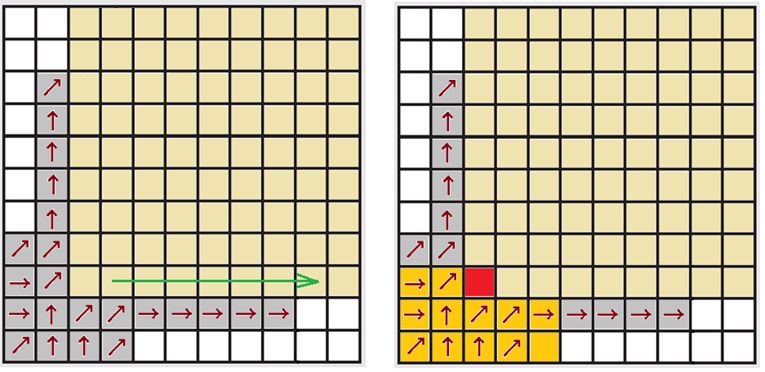}
		\caption{Initialization and order of the simulation}\label{seed}
\end{figure}

The simulation starts scanning the TVF using a fixed template to extract the pattern base. In Fig. \ref{PB} the entire pattern base using a L-form template is shown. Together with the pattern, its location $(x,y)$ in the field is also store. The simulation proceeds similar to other pattern-based methods. One starts with an empty grid, and a few rows and columns of the vector field are copied into the grid to guarantee that in the first data event all the values are known (see Fig. \ref{seed}). Gradually the grid is filled, and at each step only one grid-location is simulated following a scan order line. The template is centered in the position to be simulated, and the pattern obtained (data event), with already simulated values, is compared with the elements in the base. The pattern that is similar with the current data event must be found and the corresponding value of the chosen pattern is pasted into the position that is being simulated. The same process is repeated for each position in the grid until all the values are simulated.

To compare the two patterns, the measure $d$ is defined as considering the distance based on the TVF and the distance of their locations. It is defined as

\[ d = \beta * d_{tvf} + (1 - \beta) * d_{loc}\]

where $\beta \in [0 , 1]$. This measure is the convex combination of the distances $d_{loc}$ and $d_{tvf}$. The first is the similarity with respect to the spatial proximity between the position being simulated and the position of the pattern in the TVF. The second represents the similarity according to the vectors contained in each pattern. 

The new similarity, $d_{tvf}$, between the patterns $Pat_p$ and $Pat_q$ centered at positions $p$ and $q$, respectively, using a template $T$ is defined by
	
\begin{equation} \label{dis}
 d_{tvf} = \sum_{x\in T} (Pat_p(x) - Pat_q(x))^2
\end{equation}

In this formula, the distance $ d_{tvf}$  is only a summation of the angles' differences between two patterns. However, since not all points in the training image have an assigned direction, then it must be specified what happens when $tvf(u) =$ ND. Below is shown how the difference $Pat^{T}_p(x)-Pat^{T}_q(x)$ is calculated 

						\begin{equation*}
							\begin{aligned}
							Pat_p(x)- Pat_q(x) = 
							\begin{cases}
								tvf(x+p)-tvf(x+q), & \text{ if }Pat_p(x)\neq \text{ ND } \\ &\text{ and } Pat_q(x)\neq  \text{ ND }\\
								0, & \text{ if } Pat_p(x)= \text{ ND } \\ &\text{ and  } Pat_q(x)= \text{ ND } \\
								\frac{\pi}{b}, & \text{ otherwise }  \\
							\end{cases}
							\end{aligned}
						\label{eq:1}
						\end{equation*}

In the case when a position in one pattern has a defined direction but the corresponding position in the other pattern does not, i.e., $Pat_p(x) = $ ND and $Pat_q(x) \neq $ ND, the assigned value is $\frac{\pi}{b}$. The value of the parameter b depends on the directional interval adopted. For example, $b$ can be chosen so that $\frac{\pi}{b}$ corresponds to the length of the directional interval; this makes sense because it is the maximum value that the difference between two directions can take.    

In the pattern database $PatBase$ together with the pattern, its position was also stored. Using this position, it is defined as the local distance between the patterns $Pat_p$ and $Pat_q$ as the squared Euclidean distance of the points $p$ and $q$
\[d_{loc} = (p_x - q_ x)^2 + (p_y - q_ y)^2 \]

The value of the weight $ \beta $ in the convex combination for the measure $d$ controls the amount of non-stationarity in the simulation. If $\beta = 0 $, only the location is considered, and the modeling is not stationary, and the realization is very similar with the training directional field. When $\beta = 1 $, the pattern positions are not taken considered, and then the modeling is stationary. Hence, when a realization of the simulation is considered it will be observed to have similar features at different regions in the image. 


To compare the base of the patterns for each point, an interval $\left[0, a\right]$ is chosen (this is a parameter fixed before running the program). Then, the database is randomly traversed until it finds one pattern whose distance $d$ to the data event is in the interval. If no pattern is found at that distance, then the closest pattern is selected.  

An advantage of using the interval to select a pattern, instead of choosing the closest one, is that this introduces some degree of randomness into the program, which produces more variability in the realizations. 


A sketch of the algorithm is given in algorithm \ref{VecSim}. 

\makeatletter
\def\BState{\State\hskip-\ALG@thistlm}
\makeatother

\begin{algorithm}
\caption{VECSIM}\label{VecSim}
\begin{algorithmic}[1]
\State  A patterns base $PatBase$ is created from the TVF, using a template with an L-form. Together with each pattern its location is stored.

\State We start with an empty grid. The first $r$ rows and $t$ columns of the TVF (from left to right and bottom to top) are copied in the simulation grid; this is used as a seed.

\State Simulating the vector in the node $u$ in the grid. The simulation goes in a scan order line, starting in the left bottom corner. The template is centered at $u$ and the values are extracted, forming a pattern $P_t$. The pattern database is compared with $P_t$ using the distance

\[ d = \beta * d_{tvf} + (1 - \beta) * d_{loc}\]

\State The previous step is repeated for each node in the grid. 

\end{algorithmic}
\end{algorithm}

\section{Results}
\label{exam}

\begin{figure}[!ht]
\centering
\includegraphics[width=5.0cm]{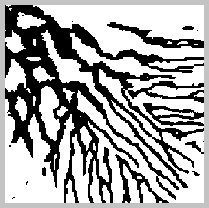}
\caption{Synthetic example of a fluvial fan-deposit with non-stationarity. \href{https://link.springer.com/journal/11004}{Reprinted by permission from Springer Nature: Springer, Mathematical Geoscience. Direct pattern-based simulation of non-stationary geostatistical models, Honarkhah M, Caers J,copyright (2012)}}\label{hona}
\end{figure}

Two examples of applications for the VECSIM method are presented, where each one is based on a different training image. In general, the training images used in MPS methods come from several different sources, such as geologist drawings, photographs, or realizations of object-based or process-based simulations. We select the first example from the reference image of \cite{HC} that corresponds to the synthetic example of a fluvial fan-deposit  with non-stationarity (see Fig. \ref{hona}). 

\begin{figure}[h!]
	  
		\centering
		\includegraphics[width=\linewidth]{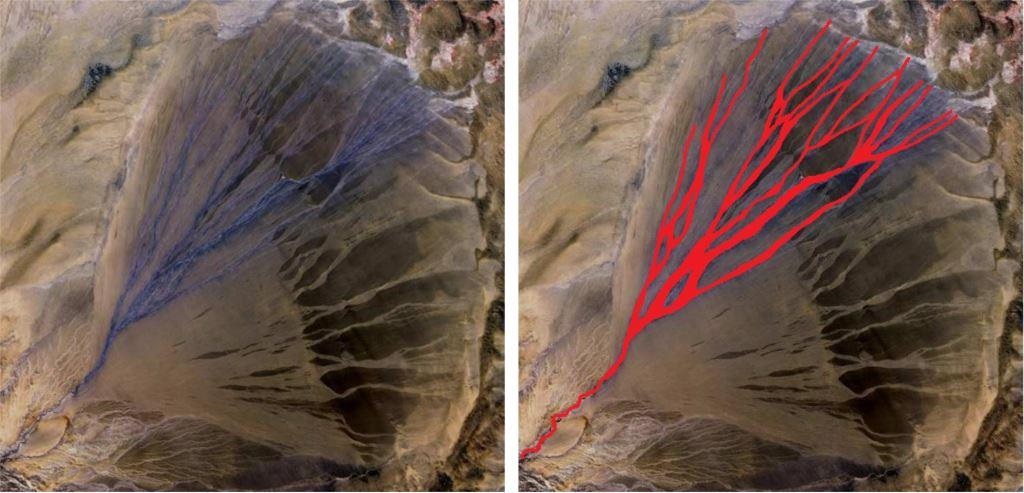}
		\caption{Right: Alluvial Fan, China. Courtesy NASA/JPL-Caltech. Left: The channel system highlighted is used as training image.}\label{china}
		       
\end{figure}

Our second training image example comes from a photograph of an alluvial fan blossoming across the landscape between the Kunlun and Altun mountain ranges that form the southern border of the Taklimakan Desert in China, given in Fig. \ref{china}. The blue region is the active part of the fan. It was highlighted in red and used as the training image  corresponding to the image 183 X 183 in the upper left corner of Fig. \ref{erosion_thin}.

For each example, first the building of the vector training fields is described and, after that, the simulation process of the image synthesis is done.


\subsection{Training vector field}
To illustrate the construction process of the vector training field, the original image with its respective erosion process, the contour curves and finally the result vector training field are presented.

\begin{figure}[h!]
	 
		\centering
		\includegraphics[width=\linewidth]{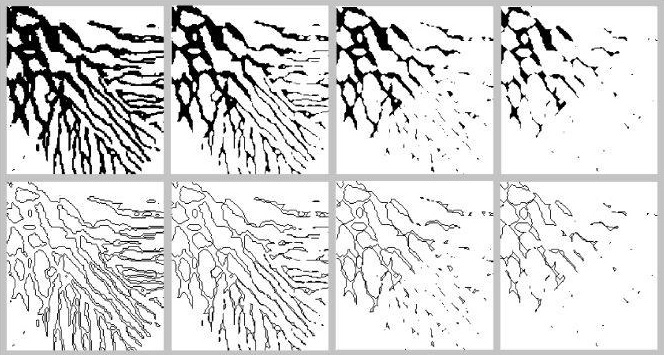}
		\caption{Erosion applied 3 times. The superior sequence presents the erosion process and inferior sequence is the corresponding contours}\label{erosion_hona}
		       
\end{figure}

Figure \ref{erosion_hona} exhibits the process of erosion, for the first example (Fig. \ref{hona}). The first line shows the images correspond to the $T_n$ sequence presented in Sect. \ref{decomposition}, respectively $T_0, T_1, T_2$ and $T_3$. In the second line, the respective contours $C_0, C_1, C_2$ and $C_3$ extracted from the corresponding image above are shown. In this example three steps of erosion are applied. 

\begin{figure}[h!]
	  
		\centering
		\includegraphics[width=\linewidth]{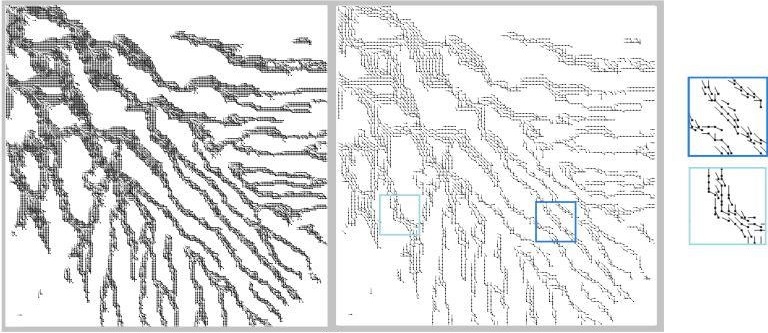}
		\caption{The training vector field and zoom in two regions}\label{tvf_hona}
		       
\end{figure}

The TVF was obtained using two steps of lengths 1 and 3. Figure \ref{tvf_hona} illustrates the training vector field obtained. On the left, the obtained field is shown at each point. On the right, the sampled points/vectors were reduced, to better reveal the structure of the vector field and to improve the visualization. 

The vector field at one point captures the local behavior around it. This is because, in part, the contour curves indicate the flow direction. Therefore, since the vectors are defined from the points of the contour, then this strategy aims to capture the direction of the flow. With this feature, we can improve the image synthesis stage by preserving the local characteristics of the model, thus keeping the important features of the original non-stationary model.

\begin{figure}[!ht]
\centering
\includegraphics[width=10.0cm]{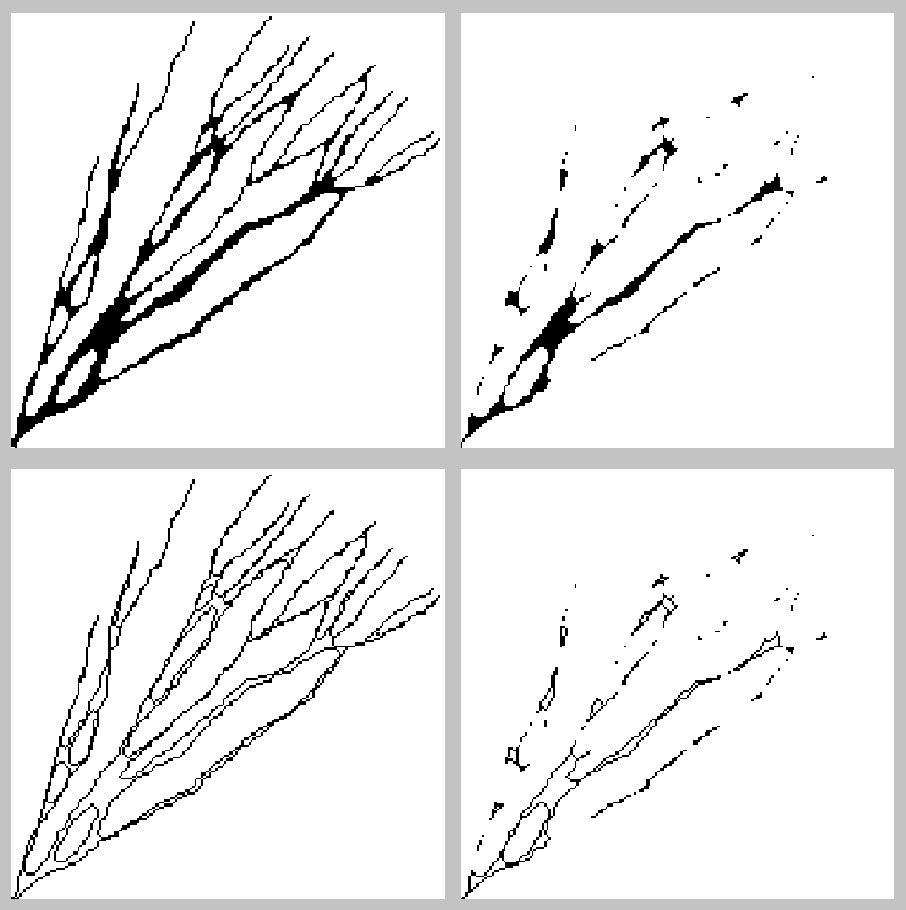}
\caption{First line exhibits erosion of the training image, second line presents the respective contours}\label{erosion_thin}
\end{figure}

In the second example of Fig. \ref{china}, the erosion process was applied 2 times in Fig. \ref{erosion_thin} . The vector was assigned to 87.4\% of the points. After that step, the interpolation was applied 6 times to designate one vector to all remaining points.

\begin{figure}[!ht]
\centering
\includegraphics[width=\linewidth]{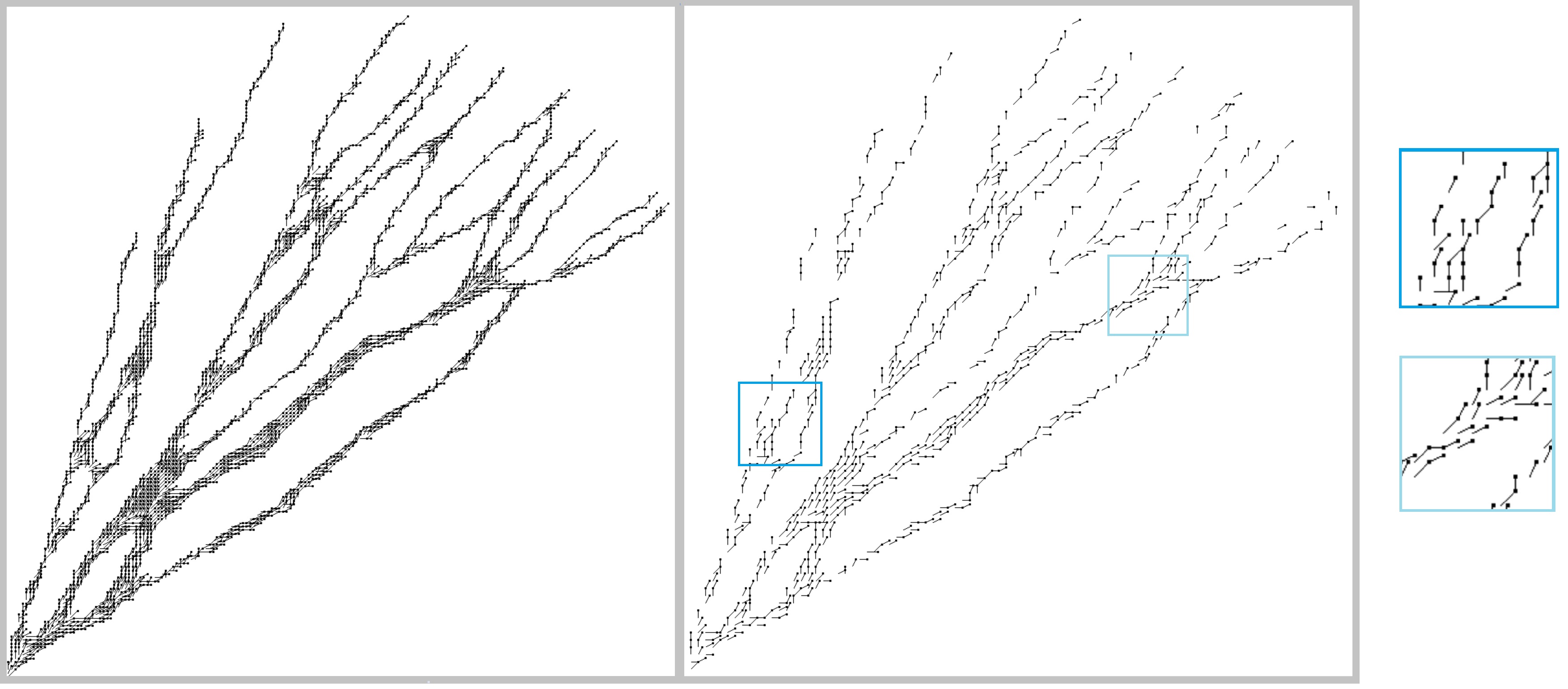}
\caption{Training vector field of training image (Fig. \ref{china}). Shown at left, the vectors generated at each point. Shown at right, the vectors sampled in selected points}\label{tvf_thin}
\end{figure}

To obtain the TVF, steps of sizes 1 and 3 were used. The directional interval assumed for this image is $ \left[0, \frac{\pi}{2}\right]$. The resulting training vector field is shown at the left of Fig. \ref{tvf_thin}. On the right, the number of vectors in the image was reduced, similar to in the first example, for a better visualization.

\subsection{Simulation}

In this section, we will show the realizations when applying VECSIM using the training vector fields previously shown.

\begin{figure}[h!]
	
		\centering
		\includegraphics[width=\linewidth]{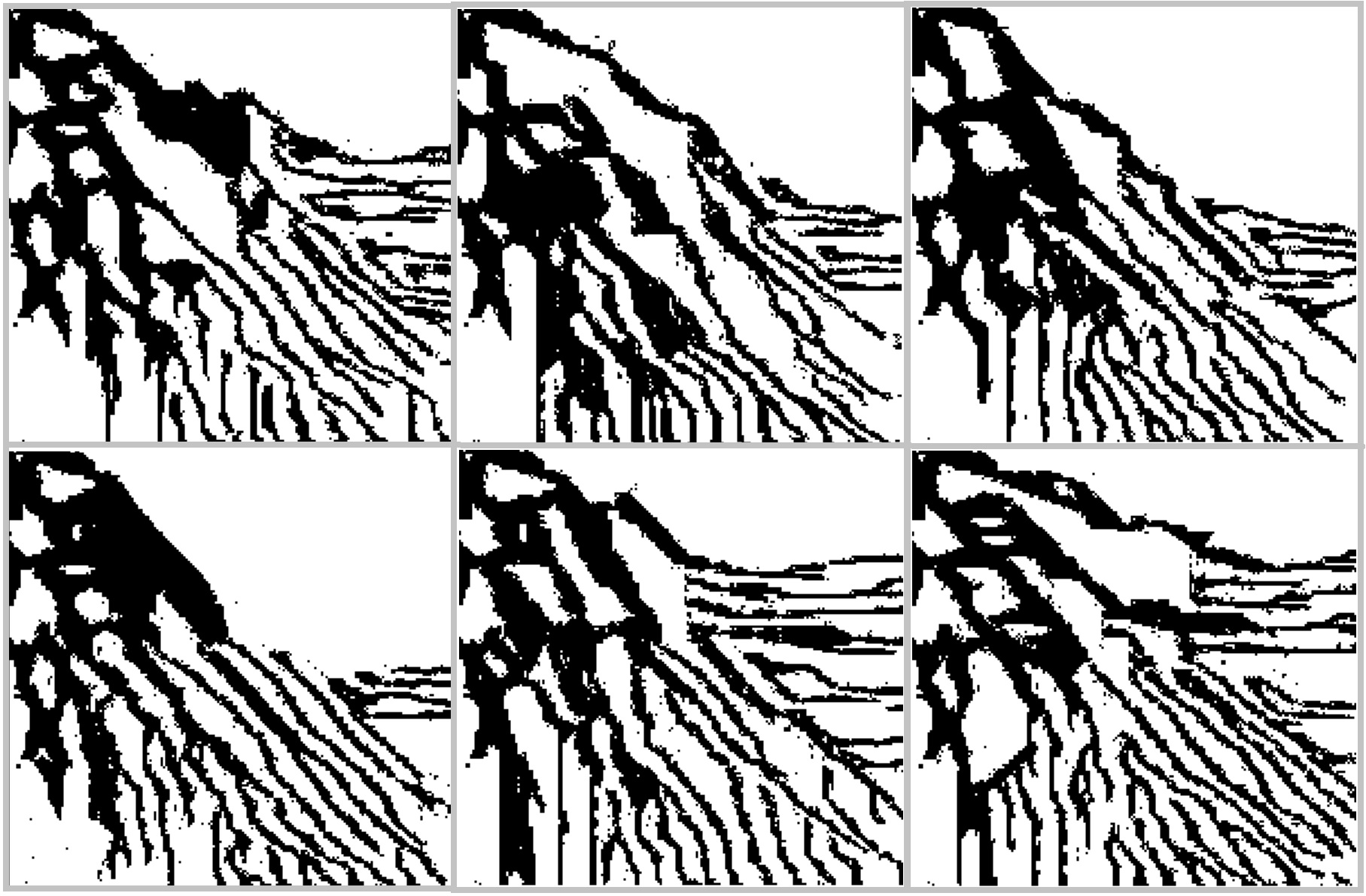}
		\caption{Realizations using the training image in Fig. \ref{hona} with $\beta $ = 0.5}\label{seq3}
\end{figure}

\begin{figure}[h!]
	 
		\centering
		\includegraphics[width=\linewidth]{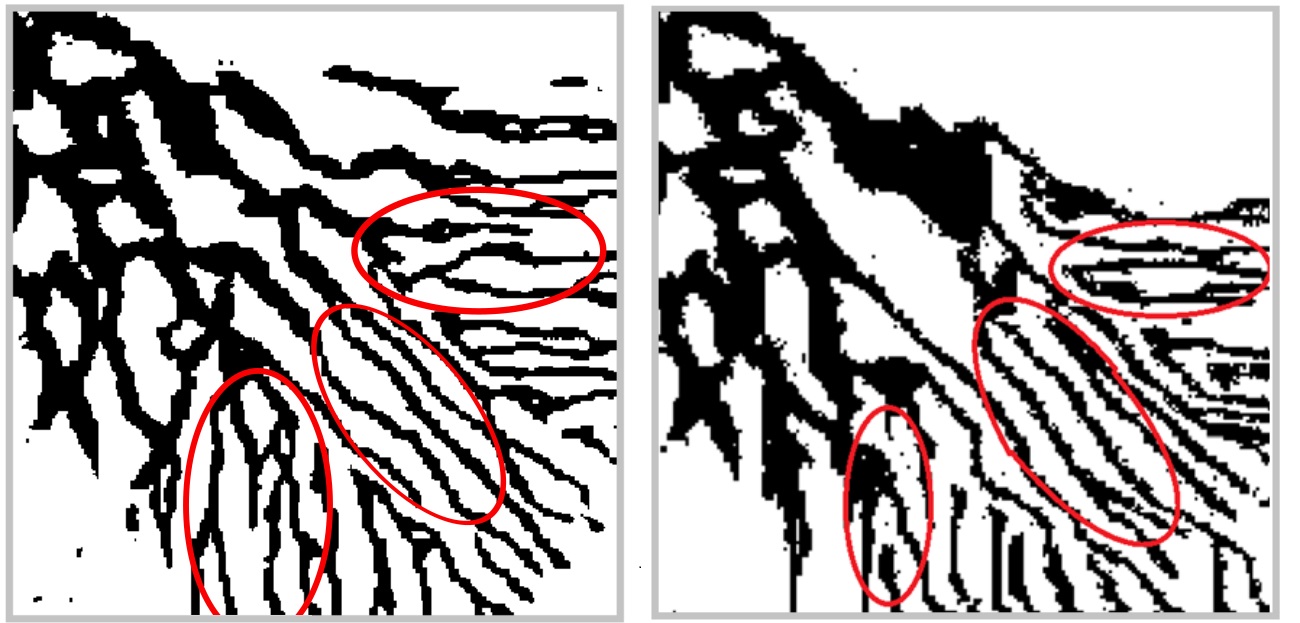}
		\caption{Left: Original training image with the three features. Right: Result of the first simulation}\label{eli}
\end{figure}

For the first example, Fig. \ref{seq3} exhibits the results of six VECSIM realizations using the training vector field given in Fig. \ref{tvf_hona} with $\beta $ = 0.5. The simulations obtained are very similar to the training image but not equal, which is desired. In the realizations shown, the method reproduces some important features of the training images. For example, in the training image three differing trends in the channel can be noted: The middle channels follow one straight line from the top left to the bottom right. On the right, the channels follow predominantly in a horizontal direction and on the left the channels follow in a vertical direction. In Fig. \ref{eli}, on the left, we indicate these three trends in the original training image. In Fig. \ref{eli} on the right, we highlight these regions in the image of the first simulation. All of the six realizations preserve a part of these features.

The trend in the channels of the training images is to start thick and finish fine. This feature is also maintained in the results. This does not mean that the simulation is very similar to the training image, however. In contrast, looking through the examples a variability between realizations can be observed. In addition, the realizations preserve the expected connectivity and the tree-like structure.

A recurrent problem when using MPS methods for elongated structures is a lack of connectivity, \cite{ML}. The algorithm proposed, VECSIM, preserves the connectivity as it appears in the training image. Normally, when the algorithm generates nonconnected images the different connected components seem to be related to the components of the training image.

\begin{figure}[h!]
\centering
\includegraphics[width=\linewidth]{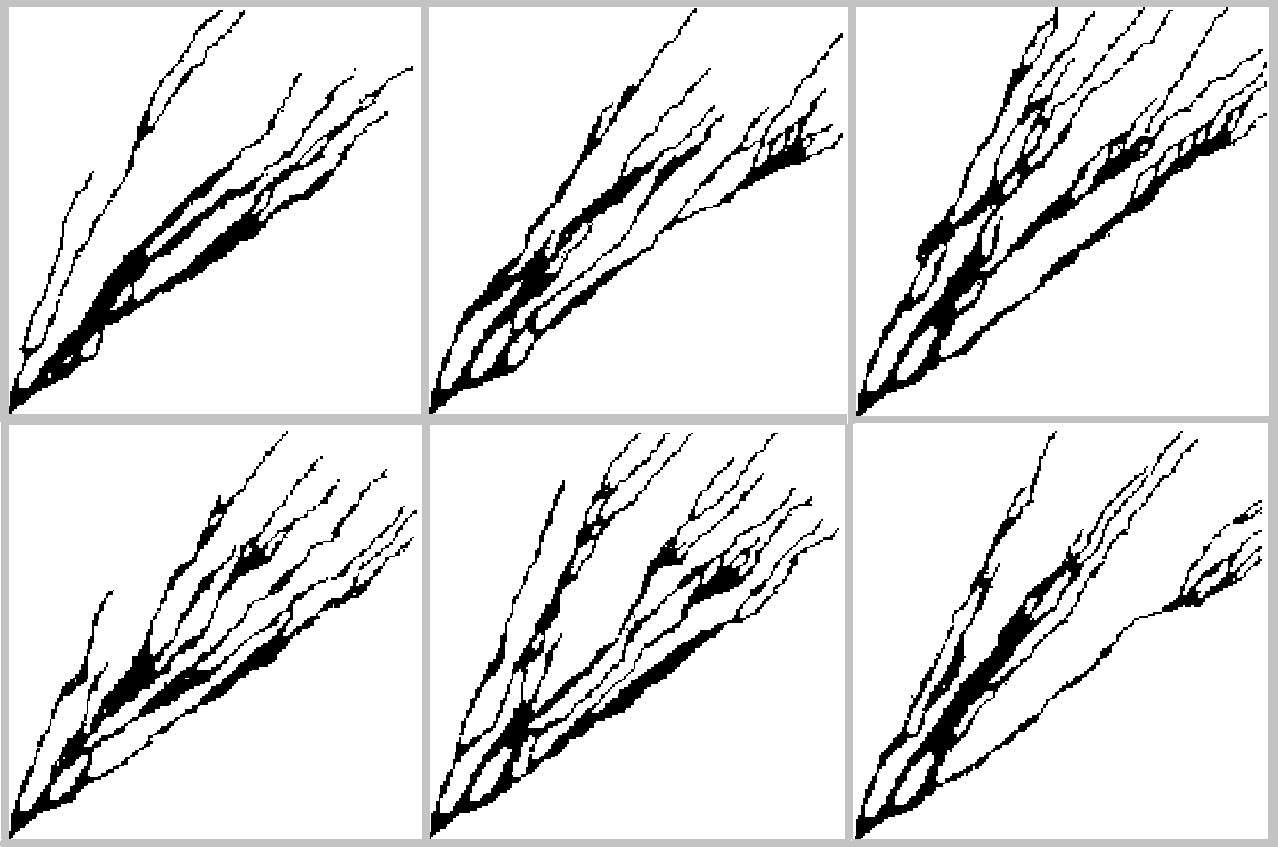}
\vspace{0.5cm}
\caption{Realizations obtained from the simulation using $\beta =0.5$} \label{realiz_nova}
\end{figure}

In the second example, Fig. \ref{realiz_nova} shows six realizations using the parameter $\beta = 0.5$. All simulations exhibit strong variability when compared with original training image. The only structure repeated in the simulations is the root of the image that was used as a seed. More importantly, the simulation has preserved important features of the training image: the distribution, the thickness and the connectivity. In the results, the channels remain well distributed inside a region limited by the leftmost and the rightmost channels and the thickness of the channels follows the training image. Finally, the training image does not have any disconnected component, and this feature is preserved in all six simulations presented.

\begin{figure}[h!]
\centering
\includegraphics[width=10.5cm]{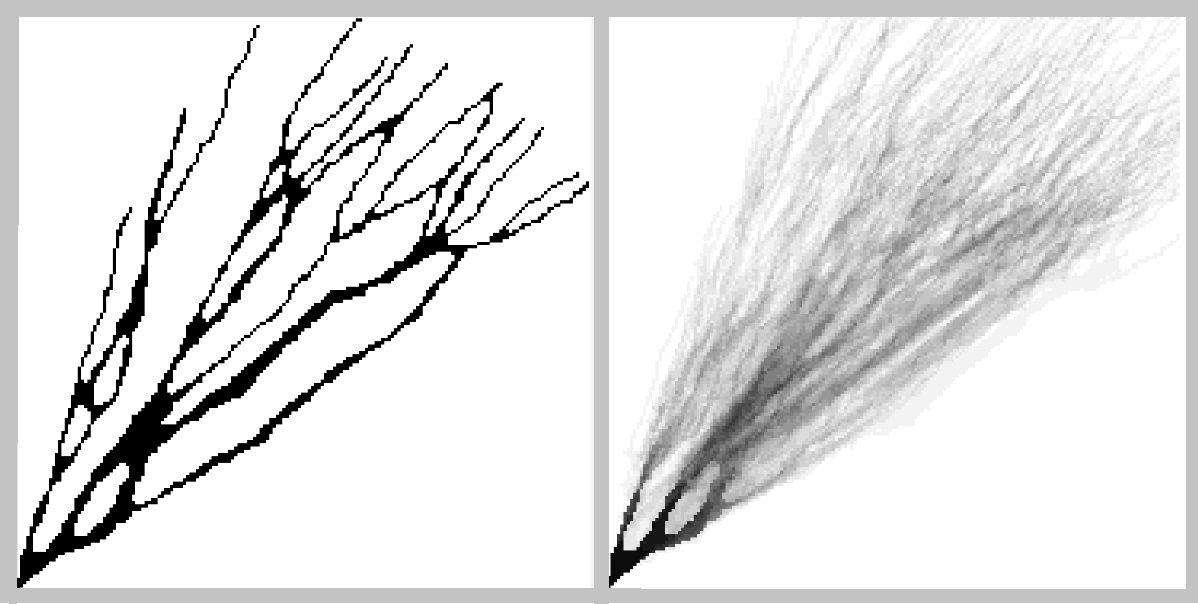}
\vspace{0.5cm}
\caption{Left: Training image. Right: E-type (average) over 30 realizations with $\beta = 0.5$}
\label{e_type}
\end{figure}

For $\beta = 0.5$, thirty realizations were generated and the E-type (ensemble averages) was computed. Figure \ref{e_type} shows the ensemble of all these realizations here can be seen that the main direction is preserved.

\section{Conclusion}
\label{conclu}

In this paper, we presented a non-stationary multiple point method using training images with a specific geometry, which was called ``tree-like''. The method is more general in nature and can be applied to images in which a vector can be defined at each point. The main contribution of this paper is the introduction of a new object called the training vector field (TVF), defined from the training image. The image being simulated should follow the flow given by the TVF. Another contribution is the definition of a new distance to compare two patterns, a definition which considers the vector information in the calculation.

The realizations of the simulation present the same geometrical characteristics as the training image, but they are not equal and the results present a strong variability. 

The similarity between patterns is computed using the position in the training image and the position simulated, so the algorithm only works when the simulation region has the same dimensions as the training image.

The algorithm proposed, VECSIM, preserves the connectivity as it appears in the examples presented. This connectivity can be compared with other works; for example, in \cite{HC}, two non-stationary simulations are presented. One of them generates an automatic segmentation of the training image in regions where a stationary modeling is applied. In the second method presented, the non-stationarity comes from the use of the pattern position in the training image of the simulation. In both cases, one can observe that the continuity presented in the training image is not preserved. The training image has some disconnected components, but in the realizations the quantity of these components is considerably higher. 

In \cite{AR}, two realizations of the method are presented. Their idea was to divide the training image into regions where a stationarity simulation is performed, and the training image used is a rotation of the original one (the rotation angle depends on the orientation presented at the region). The use of rotations explains why the channel system has a width similar in all regions. In addition to the rotation, a dilation was also employed to transform the original training image. Thus, the realization presents a width that depends on the position in the image. In these realizations, the continuity is maintained.

The use of the TVF can be introduced in many other methods for MPS and texture synthesis. Clearly, this method would only work for training images where the TVF makes sense (i.e., satisfy the tree-like condition 1). This can be thought of as passing from a $C^0$ distance to a $C^1$ distance; here, we make an analogy with the $C^0$ and $C^1$ distances for differentiable functions.

\bibliographystyle{plain}
\bibliography{Paper_v8}

\end{document}